\documentstyle[draft,aps,epsfig]{revtex}
\begin{document}
\topmargin=0.5cm
\title{Low momentum nucleon-nucleon potential \\
and  shell model effective interactions } 
\author{Scott Bogner$^{1}$, T. T. S. Kuo$^{1}$, L. Coraggio$^2$, 
A. Covello$^2$ and N. Itaco$^2$}
\address{ $^1$Department of Physics, SUNY, Stony Brook, New York 11794, USA\\
$^2$ Dipartimento di Scienze Fisiche,
 Universit\`a di Napoli Federico II and \\ 
  Istituto Nazionale di Fisica Nucleare,
 I-80126 Napoli, Italy }
\date{\today}

\maketitle
\begin{abstract}
A low momentum nucleon-nucleon (NN) potential $V_{low-k}$ is derived from 
meson exhange potentials  by integrating out
the model dependent high momentum modes of $V_{NN}$. 
The smooth and approximately unique $V_{low-k}$ is used as input 
for shell model calculations instead
of the usual Brueckner $G$ matrix.  Such an approach eliminates the nuclear 
mass dependence of the input interaction one finds in the $G$ matrix approach, 
allowing the same input interaction to be used in different nuclear regions. 
Shell model calculations
of $^{18}$O, $^{134}$Te and $^{135}$I using the {\bf same} input $V_{low-k}$ 
have been performed. For cut-off momentum $\Lambda$ in the vicinity of
2 $fm^{-1}$, our calculated low-lying spectra for these nuclei
are in good agreement with experiments, and are weakly dependent
on $\Lambda$.

\end{abstract}

\draft
A fundamental problem in nuclear physics has been the determination
of the effective nucleon-nucleon ($NN$) interaction used in the nuclear 
shell model, which has been successful in describing a variety 
of nuclear properties. There have been a number of successful
approaches \cite{ellis,covello,jiang,haxton} for  
this determination, ranging from empirical fits of experimental data, 
to deriving it microscopically from the bare $NN$ potential.  Despite 
impressive quantitative successes,   
the traditional microscopic approach suffers the fate of being 
"model dependent" owing to the fact that there is no unique $V_{NN}$ to start
from.  Moreover, as the Brueckner $G$ matrix has traditionally been the
starting point,   one obtains 
different input interactions for nuclei in different mass regions as a result
of the Pauli blocking operator. 

In this  work, we propose a different approach to shell model effective
interactions that is motivated by the recent applications
of effective field theory (EFT) and the renormalization group (RG) to low 
energy nuclear systems\cite{lepage,kaplan98,epel98,EFT}.
Our aim is to remove
some of the model dependence that arises at short distances in the
various $V_{NN}$ models, and also to eliminate the 
mass dependence one finds in the $G$ matrix approach, thus allowing the same 
interaction to be used in different nuclear regions such as
those for $^{18}$O and $^{134}$Te.  
A central theme of the RG-EFT approach 
is that physics in the infrared region is insensitive to the details of
the short distance dynamics.  One can therefore have infinitely many 
theories that differ substantially at small distances, but still give the 
same low energy physics if they possess the same symmetries and
the "correct" long-wavelength structure\cite{lepage,EFT}.  
The fact that the various meson models for $V_{NN}$ share the 
same one pion tail, but differ significantly
in how they
treat the shorter distance pieces     
illustrates this explicitly as they give the
same phase shifts and deuteron binding energy.  
In RG language, the short distance pieces of $V_{NN}$ 
are like irrelevant operators 
since their detailed form can not be resolved from low energy data. 

 Motivated by these observations, we would like to derive 
a low-momentum
NN potential $V_{low-k}$ by integrating out the high momentum components
of different models of $V_{NN}$ in the sense of the RG \cite{lepage,EFT},
and investigate its suitability of being
 used  directly as a model independent effective interaction for 
shell model calculations. We shall use in the present work the CD-Bonn NN
potential \cite{cdbonn} for $V_{NN}$. In the following, we shall first describe
our method for carrying out the high-momentum integration. Shell model
calculations for $^{18}$O, $^{134}$Te and $^{135}$I using $V_{low-k}$ will then
be performed. Our results will be discussed, especially about their
 dependence on the cut-off momentum $\Lambda$.

The first step in our approach is to integrate out the model dependent
high momentum components of $V_{NN}$.  In accordance with the general
definition of a renormalization group transformation, the decimation
must be such that low energy observables calculated in the full
theory are exactly preserved by the effective theory. 
We turn to the model space methods of nuclear structure theory 
for guidance, as there has been much work in recent years
discussing their similarity
to the Wilson RG approach\cite{haxton,bognersep,BognerSchwenk}.  While 
the technical details differ,  both approaches attempt to 
thin-out, or limit the degrees of freedom one must explicitly consider to
describe the physics in some low energy regime.  Once the relevant 
low energy modes are 
identified,  all remaining modes or states are "integrated" out.  Their
effects are then implicitly buried inside the effective interaction in 
a manner that leaves the low energy observables invariant.  
One successful model-space reduction method is 
the Kuo-Lee-Ratcliff (KLR) folded diagram theory 
\cite{klr71,ko90}.  For 
the nucleon-nucleon problem in vacuum, the RG approach simply means that the 
low momentum $T$ matrix and the deuteron binding energy calculated
from $V_{NN}$ must be reproduced by $V_{low-k}$, but with all loop
integrals cut off at some $\Lambda$.  
Therefore, we start from the half-on-shell T-matrix
\begin{equation}
  T(k',k,k^2)  
= V_{NN}(k',k)   
 + \int _0 ^{\infty} q^2 dq  V_{NN}(k',q) 
 \frac{1}{k^2-q^2 +i0^+ } T(q,k,k^2 ) .
\end{equation}
We then define an effective low-momentum T-matrix by 
\begin{equation}
  T_{low-k }(p',p,p^2)  
= V_{low-k }(p',p)   
 + \int _0 ^{\Lambda} q^2 dq  V_{low-k }(p',q) 
 \frac{1}{p^2-q^2 +i0^+ } T_{low-k} (q,p,p^2),
\end{equation}
where $\Lambda$ denotes a momentum space cut-off 
(such as $\Lambda$=2$fm^{-1}$) and $(p',p)\leq \Lambda$. 
We require the above T-matrices satisfying the condition
\begin{equation}
 T(p',p,p^2 ) = T_{low-k }(p',p, p^2 ) ;~( p',p) \leq \Lambda.
\end{equation}
The above equations define  the effective low momentum interaction
 $V_{low-k}$.  In the following, let us show that the above
equations are satisfied by the solution   
\begin{equation}
V_{low-k} = \hat{Q} - \hat{Q'} \int \hat{Q} + \hat{Q'} \int \hat{Q} \int 
\hat{Q} - \hat{Q'} \int \hat{Q} \int \hat{Q} \int \hat{Q} + ~...~~,
\end{equation}
which is  just the KLR folded-diagram effective interaction \cite{klr71,ko90}.
A preliminary account of this result has been reported as a work in progress
at a recent conference \cite{bogner01}.

 In time dependent
formulation, the T-matrix of Eq.(1) can be written as $\langle k' 
\mid VU(0,-\infty ) \mid k \rangle$, $U$ being the time evolution
operator. In this way we can readily perform a diagrammatic analysis of the
T-matrix.  A general term of it may be written as 
 $\langle k' \mid (V +V\frac{1}{e(k)}V +V\frac{1}{e(k)}V\frac{1}{e(k)}V+
\cdots) \mid k \rangle $ where $e(k)\equiv (k^2- H_o)$, $H_o$ being 
the unperrurbed Hamiltonian.
 Note that the intermediate states (represented by 1 in the numerator)
cover the entire space, and $1=P+Q$ where $P$ denotes the model space
(momentum $\leq \Lambda$) and $Q$ its complement. 
Expanding it out in terms of $P$ and $Q$, a typical term of T 
is of the form  $V\frac{Q}{e} V\frac{Q}{e}V\frac{P}{e}V\frac{Q}{e}V
\frac{P}{e}V$. 
Let us define a $\hat Q$-box as $\hat Q=V+V\frac{Q}{e}V+V\frac{Q}{e}
V\frac{Q}{e}V+ \cdot \cdot \cdot$, where all intermediate states belong
to Q. One readily sees that the T-matrix can be regrouped as a $\hat Q$-box
series, namely 
 $\langle p' \mid T \mid p \rangle = \langle p' \mid [
\hat Q+\hat Q \frac{P}{e} \hat Q
+ \hat Q \frac{P}{e} \hat Q  \frac{P}{e} \hat Q +\cdot \cdot \cdot
]\mid p \rangle$. Note that all the $\hat Q$-boxes
have the same energy variable, namely $p^2$. 

This regrouping is depicted
in Fig. 1, where each $\hat Q$-box is denoted by a circle and the solid
line represents the propagator $\frac{P}{e}$. The diagrams
A, B and C are respectively the one- and two- and three-$\hat Q$-box
terms of T, and clearly T=A+B+C+$\cdots$.
 Note the  dashed vertical line is not a propagator; 
it is just a ``ghost'' line to indicate the external indices.
 We now perform a folded-diagram factorization for the T-matrix, 
following  closely the KLR folded-diagram method 
\cite{klr71,ko90}. Diagram B of Fig. 1 is factorized into the product
of two parts (see B1) where the time integrations of the two parts
are independent from each other, each integrating from $-\infty$ to
0. In this way we have introduced a time-incorrect contribution
which must be corrected. In other words B is not equal to B1, rather
it is equal to B1 plus the folded-diagram correction B2.
 Note that the integral
sign represents a generalized folding \cite {klr71,ko90}. 

Similarly we factorize the 
three-$\hat Q$-box term C as shown in the third line of Fig. 1.
Higher-order $\hat Q$-box terms are also factorized following the same
folded-diagram procedure.
 Let us now collecting terms in the figure in a ``slanted'' way. 
The sum
of terms A1, B2, C3... is just the low-momentum 
effective interaction of Eq.(4). 
(Note that the leading $\hat Q$-box of any folded term must be at least
second order in $V_{NN}$, and hence it is denoted as  $\hat Q'$-box
which equals to $\hat Q$-box with terms first-order in $V_{NN}$ subtracted.)
The sum B1, C2, D3.... is $V_{low-k}\frac{P}{e}\hat Q$. 
Similarly the sum C1+D2+E3+$\cdots$
is just $V_{low-k}\frac{P}{e}\hat Q\frac{P}{e}\hat Q$. (Note diagrams
D1, D2, $\cdots$, E1, E2, $\cdots$ are not shown in the figure.)
Continuing this way, it is easy to see that 
 Eqs. (1) to (3) are satisfied by the low momentum  effective interaction
 of Eq.(4). 

The effective interaction of Eq.(4) can be calculated using iteration methods.
A number of such iteration methods have been developed;
the Krenciglowa-Kuo \cite{krmku74}  and the Lee-Suzuki iteration methods
\cite{suzuki80} are two examples.
These methods were formulated primarily
for the case of degenerate $PH_0P$, $H_0$ being the unperturbed
Hamiltonian.  For our present two-nucleon problem, $PH_0P$  
is obviously non-degenerate. Non-degenerate iteration methods 
 \cite{kuo95} are more complicated.
However, a recent iteration method developed by Andreozzi \cite{andre96} 
is particularly efficient for the non-degenerate case. 
This method shall be referred to as the 
Andreozzi-Lee-Suzuki (ALS) iteration method, and has been employed 
in the present work.

We have carried out numerical checks to ensure that certain low-energy
physics of $V_{NN}$ are indeed preserved by $V_{low-k}$.
We first check the deuteron binding energy $BE_d$. 
We have calculated $BE_d$ using $V_{low-k}$ for many values of $\Lambda$,
and for all cases the $BE_d$ given by $V_{low-k}$ agrees very accurately 
(to 4 places after the
decimal) with that given by $V_{NN}$. (Note that when $\Lambda$ approaches
$\infty$ $V_{low-k}$ is the same as $V_{NN}$.)
In Fig. 2, we present some $^1 S_0$
and $^3P_0$ phase shifts calculated from the CD-Bonn $V_{NN}$ (dotted line)
and the $V_{low-k}$ (circles) derived
from it, using a momentum cut-off $\Lambda =2.0 fm ^{-1}$. 
As seen, the phase shifts
 from the former are well reproduced by the latter. We have also checked the
half-on-shell T-matrix given by $V_{NN}$ and by $V_{low-k}$, and found very
good agreement between them \cite{bogner01}. The above agreements are
expected, as we have shown that the T-matrix equivalence of Eq.(3)
holds for any $\Lambda$.
In short, our numerical checks have reaffirmed
that the deuteron binding energy, low energy phase shifts and low momentum
half-on-shell T-matrix of $V_{NN}$ are all preserved by $V_{low-k}$.
As far as those physical quantities are concerned, $V_{low-k}$ and $V_{NN}$
are equivalent.



Having proven the "physical equivalence" of $V_{low-k}$ and $V_{NN}$
in the sense of the RG, we turn now to microscopic shell model calculations
in which we use $V_{low-k}$ as the input interaction.  
 A folded-diagram formulation \cite{ko90,covello,jiang} is employed.
 An important feature here is that this formalism
allows us to calculate the energy differences of neighboring many body
systems. For example, we can calculate the energy
difference of $^{18}O$ and the ground state energy of $^{16}O$,
starting from 
the experimental $^{17}O$ single particle energies (s.p.e.) and a shell model
effective interaction $V_{eff}$ derived microscopically 
from an underlying NN potential.
 This folded diagram method has been rather successfully 
applied to many nuclei using G-matrix interactions. \cite{covello,jiang}
There the basic input to the calculation are the matrix elements 
$\langle n_1 n_2 \mid
G \mid n_3 n_4 \rangle$ where $G$ is the Brueckner G-matrix and
the $n$'s are harmonic oscillator wave functions.

Since $V_{low-k}$ is already
a smooth potential, it is no longer necessary
to first calculate the G-matrix. Thus in our present work,  
the starting basic input are just the matrix elements $\langle n_1 n_2 \mid
V_{low-k} \mid n_3 n_4 \rangle$, and thereafter 
our calculation procedures are exactly
the same as described in references\cite{covello,jiang}.
A model space with two valence neutrons in the $(0d_{5/2},0d_{3/2},1s_{1/2})$ 
shell is used for $^{18}$O, and one with two and three valence protons in the 
$(0g_{7/2},1d_{5/2},1d_{3/2},2s_{1/2},0h_{11/2})$ shell for $^{134}$Te
and $^{135}$I, respectively. 
As customary, we use s.p.e. extracted from the
experimental spectra of the corresponding single-particle valence
nuclei, $^{17}$O and $^{133}$Sb \cite{nndc}.
For the absolute scaling of the sets of s.p.e., the mass-excess values
for $^{17}$O and $^{133}$Sb have been taken from
Ref. \cite{audi93,fogelberg99}. 
For $^{134}$Te and $^{135}$I, we assume that the contribution of the
Coulomb interaction between valence protons is equal to the matrix
element of the Coulomb force between the states
$(g_{\frac{7}{2}})^2_{J^{\pi}=0^+}$.

As shown in Fig. 3, our calculated low-lying $J^{\pi}$ states of $^{18}$O
 agree highly satisfactorily with experiments \cite{nndc}.
In the same figure, results of the corresponding $G$-matrix calculations
are also shown; the $V_{low-k}$ results  are just as good or slightly
better.  It may be mentioned that our $V_{low-k}$ is slightly non-hermitian. 
A hermitian $V_{low-k}$ can be obtained using the
Okubo transformation \cite{epel98}. We have constructed such
a hermitian $V_{low-k}$ using the Suzuki-Okamoto method \cite{kuoellis}. 
We have found that the shell model energy levels given by the
two $V_{low-k}$'s are very similar, probably because
 our $V_{low-k}$ is only slightly non-hermitian. 
In a concurrent paper \cite{BognerSchwenk} we have found 
that $V_{low-k}$ is almost independent of the underlying $V_{NN}$ for 
the values of $\Lambda$ considered here.  Therefore, 
although the    CD-Bonn 
potential \cite{cdbonn} is used in our present calculations, we 
stress that similar results
will be obtained if we calculate $V_{low-k}$ from other models such as 
the Paris or Argonne V-18 potentials.

  It may be mentioned that the G-matrix is energy dependent and
Pauli blocking dependent, while $V_{low-k}$ is not. This is a 
desirable feature, indicating that $V_{low-k}$ may be suitable also for 
other nuclear regions. To study this point, we have used the same
$V_{low-k}$ in a shell model calculation of $^{134}$Te as mentioned earlier.
It is encouraging that our results for $^{134}$Te also agree well
with experiments \cite{nndc} as shown in Fig. 4. 
Again the $V_{low-k}$ results are just as good or slightly
better than the G-matrix results.
We emphasize that we have used {\itshape the same\/} $V_{low-k}$ interaction 
in both
$^{18}$O and $^{134}$Te calculations, and it appears to work equally well for
both nuclei.  This is in marked contrast to the
traditional approach in which one has to use different $G$-matrices
for different mass regions, as the associated Pauli blocking
operators are different.  This is an  appealing
result, as it suggests the possibility for a common shell-model
interaction that attenuates much of the dependence on the $V_{NN}$
model and is suitable for a wide range of nuclei.

It is of  interest to investigate if the same $V_{low-k}$ is suitable
for nuclei with more than two valence nucleons. It is primarily for this
purpose we have carried out the shell model calculation of $^{135}$I
 mentioned earlier, using the same $V_{low-k}$. Our results
are shown in Fig. 5. It is gratifying that the calculated excitation
spectra are in very good agreement with experiments \cite{nndc}.
We note that the valence interaction energy for the three valence
nucleons given by our calculation
is slightly overbound, by about 0.3 MeV. This may be an indication
of  the need of a weak three-body force for this nucleus, which
has three valence nucleons . 
(Our $V_{low-k}$ is a two-body interaction.) We plan to study this 
topic in a future work.

 An important issue is what value one should use for $\Lambda$.
Guided by general EFT arguments, the minimum value for $\Lambda$ must be large
enough so that $V_{low-k}$ explicitly contains the necessary degrees of freedom
for the physical system.   For example, 
2$\pi$ exchanges are important for low energy nuclear physics,
 and to adequately include the corresponding
degree of freedom we need to have  $\Lambda_{min}$ larger than
$\sim m_{2\pi},i.e. \sim 1.4 fm^{-1}$. In fact we have found that $V_{low-k}$ 
varies strongly
with $\Lambda$ when it is smaller than that value.  
A general signal for $\Lambda_{min}$ is 
when the calculated physical quantities first become insensitive to
$\Lambda$\cite{lepage}.
  Conversely, we want $\Lambda$ to be smaller than 
the short distance scale $\Lambda_{max}$ at which the model
dependence of the different $V_{NN}$ starts to creep in\cite{lepage}.  
Systems in which these two  constraints are consistent with 
each other (i.e., $\Lambda_{min}<\Lambda_{max}$) are amenable 
to EFT-RG inspired
effective theories, as they possess a clear separation of scales between
the relevant long wavelength modes and the model dependent short distance 
structure. 
We have found \cite{BognerSchwenk} that  $\Lambda_{max}$ should
not be much greater than 2.0-2.5 fm$^{-1}$ as this is the scale after which 
$V_{low-k}$ first becomes strongly dependent on the particular $V_{NN}$
used. There is another consideration:
Most NN potentials are constructed to fit empirical phase
shifts up to $E_{lab}\approx 350$ MeV \cite{cdbonn}.  
Since $E_{lab} \le 2\hbar^2\Lambda^2/M$, M being the nucleon mass,
and one should require $V_{low-k}$ to reproduce the same 
empirical phase shifts, a choice of $\Lambda$ in the vicinity of 2 $fm^{-1}$ 
would seem to be appropriate .

Guided by the above considerations, we have used in our calculations 
two values for the momentum cut-off, namely   
$\Lambda=2.0$ and $2.2~{\rm fm}^{-1}$ as shown in Figs. 3 to 5.  
It is satisfying to see that the results 
are rather insensitive to the choice of $\Lambda$, 
in harmony with the EFT philosophy mentioned earlier. 
  Perhaps more importantly,  both are in satisfactory agreement
with experiments.


In summary, we have investigated a RG-EFT inspired approach to 
shell model calculations that is a "first step" towards a model independent
calculation that uses one common interaction over a wide range of nuclei. 
Using the KLR folded diagram approach in conjunction with 
the ALS iteration method, we have performed a RG decimation where 
 the model dependent pieces of $V_{NN}$ models are integrated out to obtain
a nearly unique low momentum potential $V_{low-k}$. 
This $V_{low-k}$  preserves the deuteron pole 
as well as the low energy phase shifts and half-on-shell $T$ matrix. 
We have used  $V_{low-k}$, which is a smooth potential, 
directly in shell model calculations of $^{18}$O, $^{134}$Te and $^{135}$I
 without first calculating the $G$ matrix.  The results are all in satisfactory
agreement with experiment, and they are insensitive to 
$\Lambda$ in the neighborhood of $\Lambda\approx 2 \; fm^{-1}$.  
We do feel that $V_{low-k}$ may become a promising and reliable
effective interaction for shell model calculations of few valence
nucleons, over a wide range of nuclear regions.



\begin{acknowledgments} We thank Prof. G.E. Brown and A. Schwenk 
for many discussions.
This work was supported in part by the U.S. DOE Grant No. DE-FG02-88ER40388, 
and the Italian Ministero dell'Universit\`a e della Ricerca Scientifica e 
Tecnologica (MURST).
\end{acknowledgments}

\begin{figure}
\caption{Folded-diagram factorization of the half-on-shell T-matrix.}
\label{fig.1}
\end{figure}

\begin{figure}
\caption{Comparison of phase shifts given by $V_{low-k}$ and $V_{NN}$.}
\label{fig.2}
\end{figure}

\begin{figure}
\caption{Low-lying states of $^{18}$O. }
\label{fig.3}
\end{figure}

\begin{figure}
\caption{Low-lying states of $^{134}$Te. }
\label{fig.4}
\end{figure}

\begin{figure}
\caption{Low-lying states of $^{135}$I. }
\label{fig.5}
\end{figure}

\end{document}